\documentclass[12pt]{article}
\usepackage{cite}

\newcommand{\citex}[1]{\textsuperscript{\citen{#1}}}
\title{Julian Schwinger: Nuclear Physics, the Radiation Laboratory,
Renormalized QED, Source Theory, and Beyond}
\author{Kimball A. Milton\footnote{K.A. Milton is 
Professor of
Physics at the University of Oklahoma.  He was a Ph.D. student of
Julian Schwinger from 1968--71, and his postdoc at UCLA for the
rest of the 1970s.  He has written a scientific biography of Schwinger,
edited two volumes of Schwinger's selected works, and co-authored
two textbooks based on Schwinger's lectures.} \\
Homer L. Dodge Department of Physics and Astronomy\\
University of Oklahoma, Norman, OK 73019 USA}
\date\today
\begin{document}

\maketitle

\begin{abstract}

Julian Schwinger's influence on twentieth century science is profound
and pervasive.  Of course, he is most famous for his renormalization
theory of quantum electrodynamics, for which he shared the Nobel Prize
with Richard Feynman and Sin-itiro Tomonaga.  But although this triumph
was undoubtedly his most heroic work, his legacy lives on chiefly through
subtle and elegant work in classical electrodynamics, quantum variational
principles, proper-time methods, quantum anomalies, dynamical mass generation,
partial symmetry, and more.  Starting as just a boy, he rapidly became the
pre-eminent nuclear physicist in the late 1930s, led the theoretical
development of radar technology at MIT during World War II, and then, soon
after the war, conquered quantum electrodynamics, and became the leading
quantum field theorist for two decades, before taking a more iconoclastic
route during his last quarter century.

Keywords: Julian Schwinger, nuclear physics, waveguides, 
quantum electrodynamics, renormalization, quantum action principle, 
source theory, axial-vector anomaly
\end{abstract}
\section{Introduction}

\setcounter{footnote}{1}

Given Julian Schwinger's commanding stature in theoretical physics 
for half a century,
it may seem puzzling why he is relatively unknown now to the educated public,
even to many younger physicists, while Feynman is a cult figure with his
photograph needing no more introduction than Einstein's.\footnote{An example
is the series of posters produced by the American Physical Society in which
the impression is given that Feynman was the chief innovator in quantum
electrodynamics.  In contradiction to this, Norman Ramsey has stated that
``it is my impression that Schwinger overwhelmingly deserved the
greatest credit for QED.  I don't think Feynman had an explanation of the
anomalous hyperfine structure before the [1948 APS] meeting.''\citex{nr}}
  This relative
obscurity is even more remarkable,
in view of the enormous number of eminent physicists, as well as other leaders
in science and industry, who received their Ph.D.'s under Schwinger's
direction, while Feynman had practically none.  
In part, the answer lies in Schwinger's
retiring nature and reserved demeanor.  Science, research and teaching, were
his life, and he detested the limelight.  Generally, he was not close to
his students, so few knew him well.  He was a gracious host and a good
conversationalist, and had a broad knowledge of many subjects, but he was
never one to initiate a relationship or flaunt his erudition.

His style of doing physics was also difficult to penetrate.  Oppenheimer
once said that others gave talks to show others how to do the calculation,
while Schwinger gave talks to show that only {\em he\/} could do it.
Although a commonly shared view, this witticism is unkind and untrue.
He was, in fact, a superb teacher, and generations of physicists, students
and faculty alike, learned physics at his feet. On the one hand he was
a formalist, inventing formalisms so powerful that they could lead, at
least in his hands, unerringly to the correct answer. He did not, therefore,
display the intuitive visualizations, for example, that Feynman commanded,
which eventually took over the popular and scientific culture.

But, more profoundly, he was a phenomenologist.  Ironically, even some of his
own students criticized him in his later years for his phenomenological
orientation, not recognizing that he had, from his earliest experiences
in nuclear physics, insisted in grounding theoretical physics in the
phenomena and data of experiment.  Isidor Rabi, who discovered Schwinger
and brought him to Columbia University, generally had a poor opinion
of theoretical physicists.  But Rabi was always very impressed with
Schwinger because in nearly every paper, he ``got the numbers out'' to
compare with experiment.  Even in his most elaborate field-theoretic
papers he was always concerned with making contact with the real
world, be it quantum electrodynamics, or strongly interacting hadrons.

Although his first, unpublished, paper, written at the age of 16, was
on the subject of the then poorly-understood quantum electrodynamics, Julian
Schwinger was almost exclusively a nuclear physicist until he joined the
Radiation Laboratory at MIT in 1943.  There, faced with critical deadlines
and the difficulty of communicating with electrical engineers, he perfected
variational techniques for solving the classical electrodynamic problems of
waveguides.  
As the War wound down, he started thinking about radiation produced
by electrons in betatrons and synchrotrons, and in so doing recognized that
the reactive and resistive portions of the electromagnetic mass of the electron
are united in a invariant structure.  Recruited by Harvard, he started teaching
there in 1946, and at first continued research in nuclear physics and
in classical diffraction.  The Shelter Island conference of 1947 changed all
that. He and Weisskopf suggested to Bethe that electrodynamic processes
were responsible for the Lamb shift, which had been known for some time as
the Pasternack effect.  Immediately, however, Schwinger saw that the most
direct consequence of quantum electrodynamics lay in the hyperfine anomaly
reported for the first time at Shelter Island.  He anticipated that the
effect was due to an induced anomalous magnetic moment of the electron.  The
actual calculation had to wait three months, while Schwinger took an extended
honeymoon, but by December 1947 Schwinger had his famous result for the
gyromagnetic ratio.
In the process he invented renormalization of mass and charge, only dimly
prefigured by Kramers.  This first formulation of QED was rather crude,
being noncovariant; to obtain the correct Lamb shift, a relativistic 
formulation
was required, which followed the next year.  A comedy of errors ensued:
Both Feynman and Schwinger made an incorrect patch between hard and soft
photon processes, and so obtained identical, but incorrect, predictions for
the Lamb shift, and the weight of their reputations delayed the publication
of the correct, if pedestrian, calculation by French and 
Weisskopf.\footnote{Schwinger later claimed that his 
first noncovariant
approach had yielded the correct result, except that he had not trusted it.}
By 1950 Schwinger had started his third reformulation of quantum 
electrodynamics, in terms of the quantum action principle.  At the same
time he wrote his remarkable paper, ``On Gauge Invariance and Vacuum
Polarization,'' formulated in a completely gauge-covariant way, 
which anticipated many later developments, including the axial vector anomaly.

His strong phenomenological bent eventually led him away from the mainstream
of physics.  Although he had given the basis for what is now called the
Standard Model of elementary particles in 1957, he never could accept
the existence of quarks because they had no independent existence outside
of hadrons.  (A secondary consideration was that quarks were invented by
Murray Gell-Mann, with whom a long-running feud had developed.)
He came to appreciate the notion of supersymmetry,
but he rejected notions of ``Grand Unification'' and of ``Superstrings''
not because of their structure but because he saw them as preposterous
speculations, based on the notion that nothing new remains to be found
in the enormous energy range from 1 TeV to $10^{19}$ GeV.  
He was sure that totally new, unexpected
phenomena were waiting just around the corner.  This seems a reasonable
view, but it resulted in a self-imposed isolation, in contrast, again, to
Feynman, who contributed mightly to the theory of partons and quantum
chromodynamics up to the end.

A complete biography of Julian Schwinger was published six
years ago.\citex{mm}  The present paper draws upon that book, as well
as on later interviews and research by the author.  Quotations of Schwinger
not otherwise attributed are based on an extended interview conducted for 
that book by my co-author Jagdish Mehra in 1988.

\section{Early Years}

Julian Schwinger was born in Manhattan, New York City, on February 12, 1918,
to rather well-off middle-class parents.  His father was a well-known
designer of women's clothes.  He had a brother Harold ten years older than
himself, whom Julian idolized as child.  Harold claimed that he taught
Julian physics until he was 13. Although Julian was recognized
as intelligent in school, everyone thought Harold was the bright one.
(Harold in fact eventually became a respected lawyer, and his mother
always considered him as the successful one, even after Julian received the
Nobel Prize.)  The Depression cost Julian's father his business, but he
was sufficiently appreciated that he was offered employment by other
designers; so the family survived, but not so comfortably as before.

The Depression 
did mean that Julian would have to rely on free education, which New York
well-provided in those days: A year or two at Townsend Harris High School,
a public preparatory school feeding into City College, where Julian
matriculated in 1933.  Julian had already discovered physics, first through
Harold's {\it Encyclopedia Britannica\/} at home, and then through the
remarkable institution of the New York Public Library.  

Larry Cranberg was a student at Townsend Harris at the same
time as Julian Schwinger.\citex{cranberg}  
They had some classes together, and both graduated
in January 1933, with a diploma that stated that graduates were entitled
to automatic entry to CCNY.  He recalled that Julian was  ``very,
very quiet.  He never gave recitations.  He sat in the last row, unsmiling
and unspeaking, and was a real loner.  But the scuttlebutt was that he was
 our star.  He very early showed promise,'' but Cranberg
saw nothing overt. ``Rumors were that he was not very good outside
math and physics, and that he was flunking German.''

Among the teachers at Townsend Harris, Cranberg particularly remembers 
Alfred Bender,\footnote{Bender was the father of physicist Carl Bender.
Carl's uncle Abram Bader was the physics teacher of Richard Feynman
at Far Rockaway High School.}  who was apparently not on the regular
faculty. Eileen Lebow, who recently wrote a
history of Townsend Harris High School,\citex{Lebow}
does not recall Bender's name.  Cranberg said that 
Bender ``fixed me on the course to be a physicist.  He was 
diligent, passionate, and meticulous in his recitations.  He was a great
guy, one of the best teachers at Townsend Harris.''  It seems
very likely that it was Bender to whom Schwinger referred to as an 
anonymous influence:
\begin{quote}
I took my first physics course in High School.  That instructor
showed unlimited patience in answering my endless questions about atomic
physics, after the class period was over.  Although I try, I cannot live up
to that lofty standard.\citex{archive}\end{quote}

At City College
Julian was reading and digesting the latest papers from Europe, and starting
to write papers with instructors who were, at the same time, graduate
students at Columbia and NYU. 
Joseph Weinberg, who went on to become a well-known relativist, was his closest
friend at City College.  Weinberg recalled his first meeting
with Julian.\citex{Weinberg} Because of his outstanding
laboratory reports, Weinberg had been granted the privilege of entering the
closed library stacks at City College.  One day he was seeking a mathematics
book,\citex{Townsend}
 which had been mentioned at the Math Club the day before, and
while he reached for it, another youngster was trying to get it.
They had both heard the talk, on functions which are continuous but
nowhere differentiable, and so they shared the book between them,
balancing the heavy volume on one knee each.  The other fellow kept finishing
 the page before Weinberg, who was a very fast reader.  Of course, his
impatient co-reader was Julian Schwinger.  Both were 15.
Weinberg mentioned that he usually spent
his time, not in the mathematics section of the library, but in the physics
section, which turned out to be Julian's base as well.  Weinberg complained
that Dirac's book on quantum mechanics\citex{DiracQM}
was very interesting and exciting,
but difficult to follow.  Julian concurred, and said it was because it was
polished too highly; he said that Dirac's original papers were much more
accessible.  Weinberg had never conceived of consulting the original
literature, so this opened a door for him.  This advice about over-refinement
Schwinger himself forgot to follow in later life.

Julian no longer had the time to spend in the
classroom attending lectures.  In physics and mathematics he was able
to skim the texts and work out the problems from first principles, 
frequently leaving the professors baffled with his original, unorthodox
solutions, but it was not so simple in history, English, and German.
City College had an enormous number of required courses then in all subjects.
His marks were not good, and he would have flunked out if the College had
 not also had a rather forgiving policy toward grades.

Joe Weinberg recalled another vivid incident.
 Among the required courses were two years of gymnasium.  One had to pass
exams in hurdling, chinning, parallel bars, and swimming.  Because
Weinberg and Julian had nearby lockers, they often fell into physics
conversations half dressed, and failed the class for lack of attendance.
Weinberg remembered seeing Julian's hurdling exam.  Julian ran up to
the bar, but came to a standstill when he was supposed to jump over
sideways.  The instructor reprimanded him, at which point Julian said,
{\it sotto voce}, ``there's not enough time to solve the equations of
motion.''

Edward Gerjuoy was another of Julian's classmates at City 
College.\citex{Gerjuoy}
``My main claim to fame is that Julian and I took the same course in
mechanics together, taught by a man named Shea, and I got an A and Julian
a B,'' because Julian did not do the work.  ``It took about a week before the 
people in the class realized we were dealing with somebody of a different
order of magnitude.'' At a time when knowledge of a bit of vector algebra was
considered commendable, ``Julian could make integrals vanish---he was very,
very impressive.  The only person in the classroom who didn't understand
this about Julian was the instructor himself.''  
``He was flunking out of City College in everything
except math and physics.  He was a phenomenon.  He didn't lead the conventional
life of a high school student before he came to City College''---unlike 
Gerjuoy and Sidney Borowitz he was not on the math team in high school
so they had not known him earlier---``when he
appeared he was just a phenomenon.''

Morton Hamermesh recalled another disastrous course.\citex{Hamermesh}
\begin{quote}We were in a
class called Modern Geometry.  It was taught by an old dodderer named
Fredrick B. Reynolds.  He was head of the math department.  He really knew
absolutely nothing.  It was amazing.  But he taught this course on Modern
Geometry.  It was a course in projective geometry from a miserable book by a
man named Graustein from Princeton, and Julian was in the class, but it was
very strange because he obviously never could get to class, at least not very
often, and he didn't own the book.  That was clear.  And every once in a while,
he'd grab me before class and ask me to show him my copy of this book and he
would skim through it fast and see what was going on.  And this fellow
Reynolds, although he was a dodderer, was a very mean character.\footnote{In
addition, he was also apparently a notorious antisemite. He used to
discourage Jewish students from studying mathematics, which worked to the
advantage of physics.\citex{gtalk}}
He used to
send people up to the board to do a problem and he was always sending 
Julian
to the board to do problems because he knew he'd never seen the course and
Julian would get up at the board, and---of course, 
projective geometry is a very
strange subject.  The problems are trivial if you think about them pictorially,
but Julian never would do them this way.  He would insist on doing them
algebraically and so he'd get up at the board at the beginning of the hour and
he'd work through the whole hour and he'd finish the thing and by that time
the course was over and anyway, Reynolds didn't understand the proof, and
that would end it for the day.\end{quote}

Sidney Borowitz, another classmate of Julian's, recalled that ``we had
the pleasure of seeing Julian attack a problem {\it de novo},  and this
used to drive Reynolds crazy.''\citex{Borowitz}

\section{Paper Number Zero}
Not only was Julian already reading the literature at City College,
he quickly started to do original research.
Julian had studied a paper by Christian M\o ller\citex{Moller}
in which he had
calculated the two-electron scattering cross section by using a retarded
interaction potential.  Of course, Schwinger read all of Dirac's papers on
quantum field theory, and was particularly impressed by the one on 
``Relativistic
Quantum Mechanics,''\citex{Dirac} ``in which Dirac went through his
attempt to recreate an electrodynamics in which the particles and light were
treated differently.''
In a paper of Dirac, Fock, and Podolsky,\citex{DFP}
\begin{quote} it was recognized that
 this was simply a unitary transformation of the Heisenberg-Pauli
theory,\citex{Heisenberg} in
which the unitary transformation was applied to the electromagnetic field.
 And I said to myself, `Why don't we apply a similar unitary transformation to
the second-quantized electron field?' I did that and worked out the lowest
approximation to the scattering amplitudes in unrelativistic notation.  It 
was a relativistic theory but it was not covariant.  That was in 1934, and
I would use it later; [the notion, called the {\em interaction 
representation,}] is always ascribed to Tomonaga, but I had done it much
earlier.\end{quote}

In deriving his result, Schwinger had to omit a term which ``represents
the infinite self-energy of the charges and must be discarded.'' This he
eventually came to see as a mistake: ``The last injunction merely parrots the
wisdom of my elders, to be later rejected, that the theory was fatally 
flawed, as
witnessed by such infinite terms, which at best, had to be discarded, or
subtracted.  Thus, the `subtraction physics' of the 1930s.''\citex{197}

This paper was never submitted to a journal, but was recently published
in a selection of Schwinger's works.\citex{js2}

\section{Columbia University}

It was Lloyd Motz, one the instructors at City College, who had learned 
about Julian from Harold, and with whom Julian was writing two papers,
who introduced him to Isidor I. Rabi.  
Then, in a conversation between Rabi and Motz
over the famous Einstein, Podolsky, and Rosen paper,\citex{epr}
which had just appeared, Julian's
voice appeared with the resolution of a difficulty through the completeness
principle, and Schwinger's career was assured.  Rabi, not without some
difficulty, had Schwinger transferred to Columbia, and by 1937 he had
7 papers published, mostly on nuclear physics,
 which constituted his Ph.D. thesis, even though his
bachelor's degree had not yet been granted.

Schwinger still
was derelict in attending classes, and ran into trouble in a chemistry
course taught by Victor LaMer. It was a dull course with a dull exam.  
A question on the final exam was ``Prove
that $d\epsilon=d\xi+d\eta$,'' where none of the variables $\epsilon$, 
$\xi$, or $\eta$ were defined.   Rabi recalled,\citex{Rabi}
\begin{quote}LaMer was, for a chemist, awfully good.  A great part of his
lifework was testing the Debye-H\"uckel theory\citex{dh}
rather brilliantly.  But he was
this rigid, reactionary type.  He had this mean way about him.  He said, `You
have this Schwinger?  He didn't pass my final exam.' I said, `He didn't?  I'll
look into it.' So I spoke to a number of people who'd taken the same course.
And they had been greatly assisted in that subject by Julian.  So I said, 
I'll fix
that guy.  We'll see what character he has.  `Now Vicky, what sort of guy are
you anyway, what are your principles?  What're you going to do about this?'
Well, he did flunk Julian, and I think it's quite a badge of distinction 
for him,
and I for one am not sorry at this point, they have this black mark on Julian's
rather elevated record.  But he did get Phi Beta Kappa as an undergraduate,
something I never managed to do.\end{quote}

 The papers which Julian wrote at Columbia were on both
theoretical and experimental physics, and Rabi prized Julian's ability
to ``get the numbers out'' to compare with experiment.
 The formal awarding of the
Ph.D. had to wait till 1939 to satisfy a University regulation.  In
the meantime, Schwinger was busy writing papers
(one, for example, was fundamental for the theory of nuclear magnetic
resonance,\citex{nmr})
and spent a somewhat lonely, but productive winter (1937) in 
Wisconsin,\footnote{It was a cold winter as well, for he failed to
unpack the trunk in which his mother had placed a warm winter coat.} where
 he laid the groundwork for his
prediction that the deuteron had an electric quadrupole moment,\citex{13}
 independently
established by his experimental colleagues at Columbia a year 
later.\citex{quad}
Wisconsin confirmed his predilection for working at night, so as not
to be ``overwhelmed'' by his hosts, Eugene Wigner and Gregory Breit.

Rabi later amusingly summarized Schwinger's year in Wisconsin.\citex{Rabi}  
\begin{quote} I
thought that he had about had everything in Columbia that we could offer---by
we, as theoretical physics is concerned, [I mean] me.  
So I got him this fellowship to
go to Wisconsin, with the general idea that there were Breit and Wigner and
they could carry on.  It was a disastrous idea in one respect, because, before
then, Julian was a regular guy.  Present in the daytime.  So I'd ask 
Julian (I'd
see him from time to time) `How are you doing?' `Oh, fine, fine.' `Getting
anything out of Breit and Wigner?' `Oh yes, they're very good, very good.' I
asked them.  They said, `We never see him.' And this is my own theory---I've
never checked it with Julian---that---there's one thing about Julian you all
know---I think he's an even more quiet man than Dirac.  He is not a fighter in
any way.  And I imagine his ideas and Wigner's and Breit's or their
personalities did not agree. I don't fault him for this, but he's such a gentle
soul, he avoided the battle by working at night.  He got this idea of working
nights---it's pure theory, it has nothing to do with the 
truth. \end{quote}

But the theory seems validated.

\section{Two Years in Berkeley}
By 1939, Rabi felt Schwinger had outgrown Columbia, so with a NRC Fellowship,
he was sent to J. Robert 
Oppenheimer in Berkeley.  This exposed him to new fields: quantum 
electrodynamics (although as we recall,
he had written an early, unpublished paper on the
subject while just 16) and cosmic-ray physics, but he mostly continued to
work on nuclear physics.  He had a number of collaborations; the most
remarkable was with William Rarita, who was on sabbatical from Brooklyn
College;  Rarita was Schwinger's ``calculating arm''\footnote{Left-leaning
Joe Weinberg accused Julian of exploiting Rarita, but Julian responded
that these papers established Rarita's reputation.}
 on a series of papers
extending the notion of nuclear tensor forces which he had conceived in
Wisconsin over a year earlier.  Rarita and Schwinger also wrote the
prescient paper on spin-3/2 particles,\citex{25}
which was to be influential decades later with the birth of supergravity.

Ed Gerjuoy, who had been an undergraduate with Schwinger at City College
in 1934,  now was one of Oppenheimer's graduate students.  He 
recalled\citex{Gerjuoy} an
amusing incident which happened one day while he, Schwinger, and Oppenheimer
were talking in Oppenheimer's office in LeConte Hall.  Two 
other students, Chaim
Richman and Bernard Peters, came in seeking a suggestion for a research problem
from Oppenheimer.  Schwinger listened with interest while Oppenheimer proposed
calculating the cross section for the electron disintegration of the deuteron.
That midnight, when Gerjuoy came to pick up Schwinger for the latter's 
breakfast
before their all-night work session, he noted that Schwinger, while waiting 
for him in the lobby of the International House, 
where Julian was living, had filled 
the backs of several telegram blanks with calculations on this problem.  
Schwinger stuffed the sheets in his pocket and they went to work.
Six months later, again Gerjuoy and Schwinger were in Oppenheimer's office when
Richman and Peters returned beaming.  They had solved the problem, and they 
covered the whole board with the elaborate solution.  Oppenheimer looked at it,
said it looked reasonable, and then asked, ``Julie, didn't you tell me you 
worked
this cross section out?''  Schwinger pulled the yellowed, crumpled blanks from
his pocket, stared at them a moment, and then pronounced the students' solution
okay apart from a factor of two.  Oppenheimer told them to find their error, 
and
they shuffled out, dispirited.  Indeed, Schwinger was right, they found they 
had
made a mistake, and they published the paper,\citex{peters}
but they were sufficiently crushed that both switched to experimental physics.

At the time, Schwinger and Gerjuoy were collaborating on a paper\citex{30} 
following from Schwinger's tensor theory of nuclear forces.  The work
\begin{quotation} \noindent
involved calculating about 200 fairly complicated spin sums,
which sums Julian and I performed independently and then compared.  To have
the privilege of working with Julian meant I had to accommodate myself to his
working habits, as follows.  Except on days when Julian had to come into the
university during conventional hours to confer with Oppenheimer, I would meet
him at 11:45 pm in the lobby of his residence, the Berkeley International
House.  He would then drive us to some Berkeley all-night bistro where he had
breakfast, after which we drove to LeConte Hall, the Berkeley physics building,
where we worked until about 4:00 am, Julian's lunchtime.  After lunch it was
back to LeConte Hall until about 8:30 am, when Julian was ready to think 
about dinner and poor TA me would meet my 9:00 am recitation class.  Since I
had just gotten married, and still was young enough to badly need my sleep,
these months of working with Julian were trying, to put it mildly.

What made it even more trying is the fact that when Julian and I carefully 
worked out together the 20 or so spin sums where our independent calculations
disagreed, Julian proved to be right every time!  I accepted the fact that 
Julian was a much better theorist than I, but I felt I was at least pretty
good, and was infuriated by his apparent constitutional inability to make a 
single
error in 200 complicated spin sum calculations.  This inability stood Schwinger
well when he embarked on the calculations that earned him the Nobel Prize.
\dots [Al]though Julian certainly realized how extraordinarily talented he
was, he never gloated about his error free calculations or in any other
way put me down.\citex{gerjouynewsletter}
\end{quotation}

The year of the NRC Fellowship was followed by a second year at Berkeley
as Oppenheimer's assistant.  They wrote an important paper together which
would prove crucial nearly a decade later:  Although Oppenheimer was
happy to imagine new interactions, Schwinger showed that an anomaly in
fluorine decay could be explained by the existence of 
{\em vacuum polarization,} 
that is, by the virtual creation of electron-positron pairs.\citex{15}
This gave
Schwinger a head start over Feynman, who for years suspected that vacuum
polarization did not exist.

\section{The War and the Radiation Laboratory}

After two years at Berkeley, Oppenheimer and Rabi arranged a real job for
Schwinger: He became first an instructor, then an Assistant Professor
at Purdue University, which had acquired a number of bright young
physicists under the leadership of Karl Lark-Horowitz.  But the war was
impinging on everyone's lives, and Schwinger was soon recruited into
the work on radar.  The move to the MIT Radiation Laboratory took place in
1943.  There Schwinger rapidly became the theoretical leader, even though
he was seldom seen, going home in the morning just as others were arriving.
He developed powerful variational methods for dealing with complicated
microwave circuits, expressing results in terms of quantities the engineers
could understand, such as impedance and admittance.

At first it seems strange that Schwinger, by 1943 the leading nuclear
theorist, should not have gone to Los Alamos, where nearly all his
colleagues eventually settled for the duration.
There seem to be at least three reasons why Schwinger stayed at the
Radiation Laboratory throughout the war.
\begin{itemize}
\item The reason he most often cited later in life was one of moral
repugnance.  When he realized the destructive power of what was being
constructed at Los Alamos, he wanted no part of it.  In contrast,
the radiation lab was developing a primarily defensive technology, radar,
which had already saved Britain.
\item He believed that the problems to solve at the Radiation Laboratory
were more interesting.  Both laboratories were involved largely in
engineering, yet although Maxwell's equations were certainly well known,
the process of applying them to waveguides required the development of
special techniques that would prove invaluable to Schwinger's later
career.
\item Another factor probably was Schwinger's fear of being overwhelmed.
In Cambridge he could live his own life, working at night when no one
was around the lab.  This privacy would have been much more difficult
to maintain in the microworld of Los Alamos.
Similarly, the working conditions at the Rad Lab were much freer
than those at Los Alamos.  Schwinger never was comfortable in a team
setting, as witness his later aversion to the atmosphere at the
Institute for Advanced Study.
\end{itemize}

 The methods and the discoveries he made at the Rad Lab
concerning the reality of the electromagnetic
mass would be invaluable for his work on quantum electrodynamics a few
years later.  As the war wound down, physicists started thinking about
new accelerators, since the pre-war cyclotrons had been defeated by
relativity, and Schwinger became a leader in this development: He proposed
a microtron,\footnote{The microtron is usually attributed to Veksler.} an 
accelerator based on acceleration through microwave cavities,
developed the theory of stability of synchrotron orbits, and most 
importantly,
worked out in detail the theory of synchrotron radiation,\footnote{This was
first circulated as a preprint in 1945.  
The paper\citex{56} published in 1949 was substantially different.}
at a time when
many thought that such radiation would be negligible because of destructive
interference.  Schwinger never properly wrote up the work he conducted in
his one and one-half years at the Rad Lab, an omission that has now be
rectified in part by publication of a volume based on his lectures then
and later, and including both published and unpublished papers.\citex{er}

Although he never really published his considerations on self-reaction,
he viewed that understanding as the most important part of his work on
synchrotron radiation:
\begin{quote}It was a useful thing for me for what was
to come later in electrodynamics, because the technique I used for calculating
the electron's classical radiation was one of self-reaction, and I did it
relativistically, and it was a situation in which I had to take seriously 
the part
of the self-reaction which was radiation, so why not take seriously the part of
the self-reaction that is mass change?  In other words, the ideas of mass
renormalization and relativistically handling them were already present at this
classical level.\end{quote}

Just after the Trinity atomic bomb test, Schwinger traveled to Los Alamos
to speak about his work on waveguides, electromagnetic radiation, and his
ideas about future accelerators.  There he met Richard Feynman for the first
time.  Feynman recalled that at the time Schwinger\citex{rpf}
\begin{quote} had already a great reputation because he had done so much work
\dots and I was very anxious to see what this man was like.  I'd always thought
he was much older than I was [they were the same age] because he had done so
much more.  At the time I hadn't done anything. \end{quote}

\setcounter{footnote}{0}
\section{QED}
In 1945 Harvard offered Schwinger an Associate Professorship,\footnote{He
beat out Hans Bethe for the job.}  which he
promptly accepted, partly because in the meantime he had met his future
wife Clarice Carrol.  Counteroffers quickly appeared, from Columbia,
Berkeley, and elsewhere, and Harvard shortly made Schwinger the youngest
full professor on the faculty to that date. There Schwinger quickly established
a pattern that was to persist for many years---he taught brilliant courses
on classical electrodynamics, nuclear physics, and quantum mechanics,
surrounded himself with a devoted coterie of graduate students and
post-doctoral assistants, and conducted incisive research that set the
tone for theoretical physics throughout the world.  

Work on classical
diffraction theory, begun at the Radiation Lab, continued for several years
largely due to the presence of Harold Levine, whom Schwinger had brought along
as an assistant.  Variational methods, perfected in the electrodynamic
waveguide context, were rapidly applied to problems in nuclear physics.
But these were old problems, and it was {\em quantum electrodynamics} that
was to define Schwinger's career.

But it took new experimental data to catalyze this development.  That
data was presented at the famous Shelter Island 
meeting held in June 1947,
a week before Schwinger's wedding.  There he, Feynman, Victor Weiss\-kopf,
Hans Bethe, and the other
participants learned the details of the new experiments of Lamb and 
Retherford\citex{lamb}
that confirmed the pre-war Pasternack effect, showing a splitting between
the $2S_{1/2}$ and $2P_{1/2}$ states of hydrogen, that should be degenerate 
according
to Dirac's theory.  In fact, on the way to the conference, Weisskopf and
Schwinger speculated that quantum electrodynamics could explain this effect,
and outlined the idea to Bethe there, who worked out the details, 
nonrelativistically, on his famous train ride to Schenectady after the 
meeting.\citex{Bethe}

But the other experiment announced there was unexpected: This was the
experiment by Rabi's group and others\citex{nafe}
of the hyperfine anomaly that would prove to
mark the existence of an anomalous magnetic moment of the electron,
expressing the coupling of the spin of the electron to an applied
magnetic field, deviating from the value again predicted by Dirac.  
Schwinger immediately
saw this as the crucial calculation to carry out first, because it
was purely relativistic, and much cleaner to understand theoretically,
not involving the complication of bound states.  However, he was delayed
three months in beginning the calculation because of an extended honeymoon
through the West.  He did return to it in October, and by December 1947
had obtained a result\citex{43}
completely consistent with
experiment.  He also saw how to compute the relativistic Lamb shift (although
he did not have the details quite right), and found the error in the
pre-war Dancoff calculation of the radiative correction to electron scattering
by a Coulomb field.\citex{Dancoff}
 In effect, he had solved all the fundamental
problems that had plagued quantum electrodynamics in the 1930s:  The
infinities were entirely isolated in quantities which renormalized the
mass and charge of the electron.  Further pro\-gress, by himself and others,
was thus a matter of technique. Concerning Schwinger's technique at the
time, Schweber writes\citex{sss}
\begin{quote}The notes of Schwinger's calculation [of the Lamb shift]
are extant [and] give
proof of his awesome computational powers. \dots These traces over photon
polarizations and integrations over photon energies \dots were carried out
fearlessly and seemingly effortlessly. \dots Often, involved steps were carried
out mentally and the answer was written down.  And, most important, the
lengthy calculations are error free!\end{quote}

\section{Covariant Quantum Electrodynamics}

During the next two years Schwinger developed two new approaches to 
quantum electrodynamics.
His original approach, which made use of successive canonical transformations,
while sufficient 
for calculating the anomalous magnetic moment of the electron, was
noncovariant, and as such, led to inconsistent results.  In particular,
the magnetic moment appeared
also as part of the Lamb shift calculation, through the coupling with the 
electric field implied by
relativistic covariance; but the noncovariant scheme gave the wrong 
coefficient. (If the coefficient
were modified by hand to the correct value, what turned out to be the correct 
relativistic value for
the Lamb shift emerged, but what that was was unknown in January 1948,
when he announced his results at the American Physical Society meeting.)

Norman Ramsey added an amusing footnote to the story about LaMer, the
chemist who flunked Julian.\citex{Ramsey}  In 1948
Schwing\-er had to repeat his brilliant lecture on quantum electrodynamics
three times at the American Physical Society meeting at Columbia, in
successively larger 
rooms.\footnote{K. K. Darrow, secretary of the
Physical Society, who apparently had little
appreciation of theory, always scheduled the theoretical sessions in
the smallest room.  Schwinger's second lecture was given in the largest
lecture hall in Pupin Lab, and the third in the largest theatre on 
campus.}
\begin{quote} It was a superb lecture.  We were impressed.  And
as we walked back together---Rabi and I were sitting together during the
lecture ---Rabi invited me to the Columbia Faculty Club for lunch.  We got in
the elevator [in the Faculty Club] when who should happen to walk in the
elevator with us but LaMer.  And as soon as Rabi saw that, a mischievous
gleam came into his eye and he
 began by saying that was the most sensational
thing that's ever happened in the American Physical Society.  The first time
there's been this three repeats---it's a marvelous revolution that's been 
done---LaMer got more and more interested and finally said, 
`Who did this marvelous thing?' And Rabi said, 
`Oh, you know him, you gave him an F, Julian Schwinger.' \end{quote}

  So first at the Pocono Conference in
April 1948,  then in the Michigan Summer School that year, and finally 
in a series of three monumental papers, 
``Quantum Electrodynamics I, II, and III,''\citex{qed123}
Julian set forth his covariant approach to QED.  At about
the same time Feynman was formulating his covariant path-integral approach; 
and although his 
presentation at Pocono was not well-received, Feynman and Schwinger compared 
notes and realized
that they had climbed the same mountain by different routes.  Feynman's 
systematic papers\citex{Feynman} were
published only after Dyson had proved the equivalence of Schwinger's and 
Feynman's schemes.\citex{Dyson}

It is worth remarking that Schwinger's approach was conservative.  
He took field theory at face value,
and followed the conventional path of Pauli, Heisenberg, and Dirac.  His 
genius was to recognize that
the well-known divergences of the theory, which had stymied all pre-war 
progress, could be
consistently isolated in renormalization of  charge and mass.  This bore a 
superficial resemblance
to the ideas of Kramers advocated as early as 1938,\citex{Kramers}
 but Kramers proceeded classically.  He had
insisted that first the classical theory had to be rendered finite and 
then quantized.  That idea was
a blind alley.  Renormalization of quantum field theory is unquestionably the 
discovery of Schwinger.

Feynman was more interested in finding an alternative to field theory, 
eliminating entirely the 
photon field in favor of action at a distance.  He was, by 1950, quite 
disappointed to realize that
his approach was entirely equivalent to the conventional electrodynamics, 
in which electron and photon fields are treated on the same footing.

As early as January 1948, when Schwinger was expounding his 
noncovariant QED to overflow crowds
at the American Physical Society meeting at Columbia University,
 he learned from Oppenheimer
of the existence of the work of Tomonaga carried out in Tokyo during the terrible conditions of
wartime.  Tomonaga had independently invented the 
``Interaction Representation'' which Schwinger
had used in his unpublished 1934 paper, and had come up with a covariant 
version of the Schr\"odinger
equation as had Schwinger, which upon its Western rediscovery was dubbed the 
Tomonaga-Schwinger equation.\citex{Tomonaga}
  Both Schwinger and Tomonaga independently wrote the same equation,
a generalization of the Schr\"odinger equation to an arbitrary spacelike
surface, using nearly the same notation.
The formalism found by Tomonaga and his school was essentially identical to 
that developed
by Schwinger five years later; yet they at the time calculated nothing, 
nor did they discover 
renormalization.  That was certainly no reflection on the ability of 
the Japanese; Schwinger could not
have carried the formalism to its logical conclusion without the impetus 
of the postwar experiments,
which overcame prewar paralysis by showing that the quantum corrections 
``were neither infinite nor zero, but finite and small, and 
demanded understanding.''\citex{197}

However, at first Schwinger's covariant
calculation of the Lamb shift contained another error, the same as
Feynman's.\citex{Feynman2} 
\begin{quote} By this time I had forgotten the number
I had gotten by just artificially changing the wrong spin-orbit coupling.
Because I was now thoroughly involved with the covariant calculation and it
was the covariant calculation that betrayed me, because something went wrong
there as well.  That was a human error of stupidity.\end{quote}  French and 
Weisskopf\citex{French} had gotten the right answer, 
\begin{quote} because they
put in the correct value of the magnetic moment and used it all the way 
through.  I, at an earlier stage, had done that, in effect, and also got the
same answer.\end{quote}

  But now he and Feynman ``fell into the same trap.  We were
connecting a relativistic calculation of high energy effects with a
nonrelativistic calculation of low energy effects, a la Bethe.''  Based
on the result Schwinger had presented at the APS meeting in January 1948,
Schwinger claimed priority for the Lamb shift calculation: 
\begin{quote} I had the
answer in December of 1947.  If you look at those [other] papers you will
find that on the critical issue of the spin-orbit coupling, they appeal
to the magnetic moment.  The deficiency in the calculation I did [in 1947]
was [that it was] a non-covariant calculation.  French and Weisskopf
were certainly doing a non-covariant calculation.  Willis 
Lamb\citex{Kroll} was doing
a non-covariant calculation.  They could not possibly have avoided these
same problems.\end{quote}

 The error Feynman and Schwinger
made had to do with the infrared problem that occurred in the relativistic
calculation, which was handled by giving the photon a fictitious mass.
\begin{quote}
Nobody thought that if you give the photon a finite mass it will also affect
the low energy problem.  There are no longer the two transverse degrees
of freedom of a massless photon, there's also a longitudinal degree of
freedom.  I suddenly realized this absolutely stupid error, that a photon
of finite mass is a spin one particle, not a helicity one 
particle.\end{quote}  Feynman was more forthright and apologetic in 
acknowledging\citex{Feynman2} his
error which substantially delayed the publication of the French and Weisskopf
paper.

\section{Quantum Action Principle}

Schwinger learned from his competitors, particularly Feynman and Dyson.  
Just as Feynman had
borrowed the idea that henceforward would go by the name of Feynman 
parameters from Schwinger,
Schwinger recognized that the systematic approach of Dyson and Feynman 
was superior in higher orders.  So by 1949 he replaced 
the Tomonaga-Schwinger approach by a much more powerful engine,
the quantum action principle.  This was a logical outgrowth of the 
formulation of Dirac,\citex{Dirac33} as was
Feynman's path integrals; the latter was an integral approach, 
Schwinger's a differential.  The
formal solution of Schwinger's differential equations was Feynman's 
functional integral; yet while the
latter was ill-defined, the former could be given a precise meaning, and 
for example, required the
introduction of fermionic variables, which initially gave Feynman some 
difficulty.  

It may be fair to say that while the path integral formulation to 
quantum field 
theory receives all the press, the most precise exegesis of field theory 
is provided by the functional differential
equations of Schwinger resulting from his action principle.
He began in the ``Theory of Quantized Fields I''\citex{65} by introducing 
a complete set of eigenvectors ``specified by a spacelike 
surface \dots and the eigenvalues \dots of a complete set of 
commuting operators constructed from field quantities attached to that
surface.''  The question is how to compute the transformation function
from one spacelike surface to another.
After remarking that this development, time-evolution,
must be described by a unitary transformation, he {\it assumed\/} that
any infinitesimal change in the transformation function must be given
in terms of the infinitesimal change in a quantum action operator, 
or of a quantum Lagrange function.  This is the quantum dynamical
principle, a generalization of the principle of least action, or of
Hamilton's principle in classical mechanics.
If the parameters of the system are not altered, the only changes
arise from those of the initial and final states, 
from which Schwinger deduced the {\it Principle of
Stationary Action},
from which the field equations may be derived.  A series of six papers followed
with the same title, and the most important ``Green's Functions of
Quantized Fields'' published in the Proceedings of the National 
Academy.\citex{66}

Paul Martin presented an entertaining account of the prehistory of their
work together.\citex{martin}
\begin{quotation}
During the late 1940s and early 1950s Harvard was the home of a
school of physics with a special outlook and a distinctive
set of rituals.  Somewhat before noon three times each week, the master
would arrive in his blue chariot and, in forceful and beautiful lectures,
reveal profound truths to his Cantabridgian followers, Harvard and M.I.T.
students and faculty.\footnote{In a later 
recollection,\textsuperscript{\citen{martin2}} Martin elaborated: ``Speaking
eloquently, without notes, and writing with both hands, he expressed what was
already known in new, unified ways, incorporating original examples and
results almost every day.  Interrupting the flow with questions was like
interrupting a theatrical performance.  The lectures continued through
Harvard's reading period and then the examination period.  In one course
we attended, he presented the last lecture---a novel calculation of the Lamb
Shift---during Commencement Week.  The audience continued coming and he 
continued lecturing.''}
 Cast in a language more powerful and general than any of
his listeners had ever encountered, these ceremonial gatherings had some
sacrificial overtones---interruptions were discouraged and since the sermons 
usually lasted past the lunch hour, fasting was often required.  Following
a mid-afternoon break, private audiences with the master were permitted and,
in uncertain anticipation, students would gather in long lines to seek
counsel.

During this period the religion had its own golden rule---the action 
principle---and its own cryptic testament---`On the Green's Functions of
Quantized Fields.'\citex{66}  Mastery of this paper conferred on followers a 
high
priest status.\footnote{Schwinger evidently was aware of the mystique.
In a later letter recommending Martin for a permanent appointment at Harvard
he stated that Martin was ``superior in intrinsic ability and performance.
Quantum field theory is the new religion of physics, and Paul C. Martin 
is one of its high priests.''\citex{archive} However, as the last
paragraph of the present essay demonstrates, Schwinger throughout his
life maintained a tension between an elitist and a democratic view of science.}
  The testament was couched in terms that could not be
questioned, in a language whose elements were the values of real physical
observables and their correlations.  The language was enlightening, but
the lectures were exciting because they were more than metaphysical.  Along
with structural insights, succinct and implicit self-consistent methods for
generating true statements were revealed.
\end{quotation}

Recently, a perceptive analysis of Schwinger's Green's functions papers
has been given by Schweber\citex{ssspnas}.  There he concludes that
\begin{quote} Schwinger's formulation of relativistic QFTs [quantum field
theories] in terms of Green's functions was a major advance in theoretical
physics.  It was a representation in terms of elements (the Green's
functions) that were intimately related to real physical observables and
their correlation.  It gave deep structural insights into QFTs; in particular,
it allowed the investigation of the structure of the Green's functions
when their variables are analytically continued to complex values, thus
establishing deep connections to statistical mechanics.
\end{quote}

\section{``Gauge Invariance and Vacuum Polarization''}

The paper ``On Gauge Invariance and Vacuum Polarization''\citex{64}, 
submitted by Schwin\-ger
 to the {\it Physical Review\/} near the end of December 1950,
is nearly universally acclaimed as his greatest publication.  As his lectures
have rightfully been compared to the works of Mozart, so this might be
compared to a mighty construction of Beethoven, the 3rd Symphony, the {\it
Eroica}, perhaps.  It is most remarkable because it stands in splendid
isolation.  It was written over a year 
after the last of his series of papers on
his second, covariant, formulation of quantum electrodynamics
was completed: ``Quantum Electrodynamics III. The Electromagnetic Properties
of the Electron---Radiative Corrections to Scattering''\citex{qed123} 
was submitted in May 1949.  And barely two months later, in March 1951,
 Schwinger would submit the first
of the series on his third reformulation of quantum 
field theory, that based on the quantum action principle, namely, 
``The Theory of Quantized Fields I.''\citex{65}
But ``Gauge Invariance and Vacuum Polarization''  stands on its own,
and has endued the rapid changes in tastes and developments in quantum
field theory, while the papers in the other series are mostly of historical
interest now.  As Lowell Brown\citex{brown}
pointed out, ``Gauge Invariance and Vacuum Polarization''
still has over one hundred citations per year, and is far and away Schwinger's 
most cited paper.\footnote{In the 2005 {\it Science 
Citation Index}, it had 
105 citations, out of a total of 458 citations to all of Schwinger's 
work.\citex{sci} These numbers have remained remarkably constant over the years.}  
Yet even such a masterpiece was not without its critics.  Abraham Klein, who
was finishing his thesis at the time under Schwinger's direction, 
and would go on to be one of Schwinger's second set of ``assistants'' 
(with Robert
Karplus), as, first, an instructor, and then a Junior Fellow, recalled that 
Schwinger (and, independently, he and Karplus) ran afoul of a temporary editor 
at the {\it Physical Review}. That editor thought Schwinger's original paper 
repeated too many complicated expressions and that symbols should be introduced
to represent expressions that appeared more than once.  Schwinger complied,
but had his assistants do the dirty work.  Harold Levine, who was still
sharing Schwinger's office, working on the never-to-be-completed 
waveguide book, typed the revised manuscript, while
Klein wrote in the many equations.  Klein recalled that he took much more
care in writing those equations than he did in his own
 papers.\citex{Klein}

Schwinger recalled later that he viewed this paper, in part, as a reaction
to the  ``invariant regularization'' of Pauli and 
Villars.\citex{pv}  \begin{quote} It was this paper,
with its mathematical manipulation, without physical insight particularly
about questions such as photon mass and so forth, which was the direct
inspiration for `Gauge Invariance and Vacuum 
Polarization.' The whole point is that if
you have a propagation function, it has a certain singularity when the two
points coincide.  Suppose you pretend that there are several particles
of the same type with different masses and with coupling constants which
can suddenly become negative instead of positive.  Then, of course, you
can cancel them.  It's cancellation again, subtraction physics, done in a
more sophisticated way, but still, things must be made to add up to zero.
Who needs it?\end{quote}

In this paper, Schwinger obtained a closed form for the electron propagator
in an external magnetic field, by solving proper-time equations of motion,
opening a field which would be fashionable nearly three decades later with
the discovery of pulsars; gave the definitive derivation of the 
Euler-Heisenberg
Lagrangian describing the scattering of light by light, a phenomenon still not
observed directly; and gave the precise connection between axial-vector and
pseudoscalar meson theories, what became known as the axial-vector anomaly when
it was rediscovered nearly two decades later by Adler, Bell, and 
Jackiw.\citex{abj}  (We will discuss this the anomaly later in  
Sec.~\ref{avanomaly}.)
The paper is not only a thing of great beauty, but a powerful storehouse of 
practical technique for solving gauge-theory problems in a gauge-invariant way.

\section{Harvard and Schwinger}

So it was no surprise that in the late 1940s and early 1950s 
Harvard was the center of the world, as
far as theoretical physics was concerned.  Everyone, students and 
professors alike, flocked to
Schwinger's lectures.  Everything was revealed, long before publication; 
and not infrequently
others received the credit because of   Schwinger's reluctance to publish 
before the subject was ripe.
A case in point is the so-called Bethe-Salpeter equation,\citex{bs}
 which as Gell-Mann and Low noted,\citex{gml} ``first
appeared in Schwinger's lectures at Harvard.''  At any one time, Schwinger 
had ten or twelve Ph.D.
students, who typically saw him but rarely.  In part, this was because he was 
available to see his
large flock but one afternoon a week, but most saw him only when absolutely 
necessary, because they recognized 
that his time was too valuable to be wasted on trivial matters.  A student 
may have
seen him only a handful of times in his graduate career, but that was all the 
student required.
When admitted to his sanctum, students were never rushed, were listened to 
with respect, treated with kindness,
and given inspiration and practical advice.  One must remember that the 
student's problems
were typically quite unrelated to what Schwinger himself was working on at 
the time; yet in a few
moments, he could come up with amazing insights that would keep the student 
going for weeks,
if not months.  A few students got to know Schwinger fairly well, and were 
invited to the Schwingers'
house occasionally; but most saw Schwinger primarily as a virtuoso in the 
lecture hall, and now and
then in his office.  A few faculty members were a bit more intimate, but 
essentially Schwinger was a very private person.

\section{Custodian of Field Theory}

Feynman left the field of quantum electrodynamics in 1950, regarding it as 
essentially complete.
Schwinger never did.  During the next fifteen years, he continued to explore 
quantum field theory,
trying to make it reveal the secrets of the weak and strong interactions.  
And he accomplished much.
In studying the relativistic structure of the theory, he recognized that all 
the physically significant
representations of the Lorentz group were those that could be derived from 
the ``attached'' four-dimensional
Euclidean group, which is obtained by letting the time coordinate become 
imaginary.\citex{86} This idea was
originally ridiculed by Pauli, but it was to prove a most fruitful suggestion. 
Related to this was the
CPT theorem, first given a proof for interacting systems by Schwinger in his 
``Quantized Field'' papers
of the early 1950s, and elaborated later in the decade.\citex{84}

 By the end of the 1950s, Schwinger, with his
former student Paul Martin, was applying field theory methods of many-body 
systems, which led to a revolution in that field.\citex{89} 
Methods suitable for describing systems out of equilibrium, usually associated
with the name of Keldysh,\citex{keldysh} were obtained some four
years earlier by Schwinger.\citex{101}
Along the way, in what he considered rather 
modest papers, he discovered Schwinger terms,\citex{90}
anomalies in the commutation relations between field operators, and the 
Schwinger model,\citex{108}
still the only known example of dynamical mass generation.  
The beginnings of a quantum field
theory for non-Abelian fields was made;\citex{105}
 the original example of a non-Abelian field being that of
the gravitational field, he laid the groundwork for later canonical 
formulations of gravity.\citex{adm}

\section{Measurement Algebra}

In 1950 or so, as we mentioned, Schwinger developed his action principle, 
which applies to any
quantum system, including nonrelativistic quantum mechanics.  Two years later, 
he reformulated
quantum kinematics, introducing symbols that abstracted the essential elements
of realistic measurements.
This was measurement algebra, which yielded conventional Dirac quantum 
mechanics.  But although
the result was as expected, Schwinger saw the approach as of great value 
pedagogically, and as
providing a interpretation of quantum mechanics that was self-consistent.  
He taught quantum mechanics
this way for many years, starting in 1952 at the Les Houches summer school; 
but only in 1959 did he
start writing a series of papers expounding the method to the world.  He 
always intended to write a
definitive textbook on the subject, but only an incomplete version based on 
the Les Houches lectures ever appeared during his lifetime.\citex{152}  Englert
has now put his later undergraduate UCLA lectures together in a lovely book
published by Springer.\citex{englert}

One cannot conclude a retrospective of Schwinger's work without mentioning
two other magnificent achievements in the quantum mechanical domain.
 He presented a definitive development
of angular momentum theory derived in terms of oscillator variables in
``On Angular Momentum,'' which was never properly
published;\citex{69}\footnote{This and other of Schwinger's most
important papers were reprinted in two selections of his work.\citex{js1,js2}}
and he developed a ``time-cycle'' method of calculating
matrix elements without having to find all the wavefunctions in his
beautiful ``Brownian Motion of a Quantum Oscillator,''\citex{101} which as
we mentioned above anticipated the work of Keldysh.\citex{keldysh}
We should also mention the famous  Lippman-Schwinger paper,\citex{60}
which is chiefly remembered for what Schwinger considered a standard
exposition of quantum scattering theory, not for the variational methods
expounded there.

\section{Electroweak Synthesis}

In spite of his awesome ability to make formalism work for him, Schwinger 
was at heart a
phenomenologist.  He was active in the search for higher symmetry; while he 
came up with $W_3$,
Gell-Mann found the correct approximate symmetry of hadronic states, $SU(3)$.  
Schwinger's
greatest success in this period was contained in his masterpiece, his 1957 
paper ``A Theory of
the Fundamental Interactions''.\citex{82} 
Along with many other insights, such as the existence of two neutrinos and
the $V-A$ structure of weak interactions,
 Schwinger there laid the
groundwork for the electroweak unification.  He introduced two charged 
intermediate vector
bosons as partners to the photon, which couple to charged weak currents. 

 A few years later,
his former student, Sheldon Glashow, as an outgrowth of his thesis, would 
introduce a neutral
heavy boson to close the system to the modern $SU(2)\times  U(1)$ symmetry 
group;\citex{Glashow}  Steven Weinberg\citex{SWeinberg}
 would complete the picture by generating the masses for the heavy bosons by
spontaneous symmetry breaking.  Schwinger did not have the details right in 
1957, in 
particular because experiment then
seemed to disfavor the $V-A$ theory his approach implied,
but there is no doubt that Schwinger must be counted as the grandfather of the 
Standard Model on the basis on this paper.

\section{The Nobel Prize and Reaction}

Recognition of Schwinger's enormous contributions had come early.  
He received the Charles L. Meyer
Nature of Light Award in 1949 on the basis of the partly completed manuscripts 
of his ``Quantum
Electrodynamics'' papers.  The first Einstein prize was awarded to him, 
along with Kurt G\"odel,
in 1951.  The National Medal of Science was presented to him by President 
Johnson in 1964, and, of
course, the Nobel Prize was received by him, Tomonaga, and Feynman from the 
King of Sweden in 1965.

But by that point his extraordinary command of the machinery of quantum field 
theory had convinced
him that it was too elaborate to describe the real world, at least directly.  
In his Nobel Lecture,
he appealed for a phenomenological field theory that would make immediate
contact with the particles experiencing
the strong interaction.  Within a year, he developed such a theory, 
Source Theory.

\section{Source Theory and UCLA}

It surely was the difficulty of incorporating strong interactions into
field theory that led to ``Particles and Sources,'' received
by the {\it Physical Review\/} barely six months after his Nobel lecture,
in July 1966,\citex{135} based on lectures Schwinger
gave in Tokyo that summer.  One must appreciate the milieu in which
Schwinger worked in 1966.  For more than
a decade he and his students had been nearly the only exponents of field
theory, as the community sought to understand weak and strong interactions,
and the proliferation of ``elementary particles,'' 
through dispersion relations,
Regge poles, current algebra, and the like, most ambitiously through the
$S$-matrix bootstrap hypothesis of Geoffrey Chew and
Stanley Mandelstam.\citex{chew,frautschi,adler,mandelstam}
 What work in field theory did exist then was largely axiomatic,
an attempt to turn the structure of the theory into a branch of mathematics,
starting with  Arthur Wightman,\citex{wightman}
and carried on by many others, including
Arthur Jaffe at Harvard.\citex{jaffe}
(The name changed from axiomatic field theory to constructive field theory
along the way.) Schwinger looked
on all of this with considerable distaste; not that he did not appreciate
many of the contributions these techniques offered in specific contexts,
but he could not see how they could form the {\em basis\/} of a theory.

The new source theory was supposed to supersede field theory, much
as Schwinger's
successive covariant formulations of quantum electrodynamics had replaced
his earlier schemes.  In fact, the revolution was to be more profound,
because there were no divergences, and no renormalization.
\begin{quote}
The concept of renormalization is simply foreign to this phenomenological
theory.  In source theory, we begin by hypothesis with the description of
the actual particles, while renormalization is a field theory concept
in which you begin with the more fundamental operators, which are then
modified by dynamics.  I emphasize that there never can be divergences
in a phenomenological theory.  What one means by that is that one is
recognizing that all further phenomena are consequences of one
phenomenological constant, namely the basic charge unit, which describes the
probability of emitting a photon relative to the emission of an electron.
When one says that there are no divergences one means that it is not
necessary to introduce any new phenomenological constant.  All further
processes as computed in terms of this primitive interaction automatically
emerge to be finite, and in agreement with those which historically had
evolved much earlier.\citex{139a}\end{quote}

\subsection{Engineering Approach to Particle Theory}

In 1969 Schwinger gave the Stanley H. Klosk lecture to the New York University
School of Engineering Science.  Because that lecture captures his
philosophy underpinning source theory so well, at an early stage in the
development of that approach, I quote the transcription
of it in full.\citex{sciresearch}

\begin{quotation}
It is a familiar situation in physics that when an experimental domain is to be
codified, even though a fundamental theory may be available, rarely is it 
brought directly to bear upon the experimental material.
The fundamental theory is too complicated, generally too remote from the 
phenomena that you want to describe.  Instead, there is always an intermediate
theory, a phenomenological theory, which is designed to deal directly with
the phenomena, and therefore makes use of the language of observation.  On the
other hand, it is a genuine theory, and employs abstract concepts that can make
contact with the fundamental theory.

The true role of fundamental theory is not to confront the raw data, but to 
explain the relatively few parameters of the phenomenological 
theory in terms of which the great mass of raw data has been organized.

I learned this lesson 25 years ago during World War II, when I became interested
in the problems of microwave systems, wave guides in particular.  Being very
naive, I started out solving Maxwell's equations.  I soon learned better.
Most of the information in Maxwell's equations is really superfluous.  As far
as any particular problem is concerned, one is only interested in the 
propagation of just a few modes of the wave guide.  A limited number of
quantities that can be measured or calculated tell you how these few modes
behave and exactly what the system is doing.

You are led directly to a phenomenological theory of the kind engineers
invariably use---a picture, say, in terms of equivalent transmission lines.  
The only role of Maxwell's equations is to calculate the few parameters, 
the effective lumped constants that characterize the equivalent circuits.

The engineer's intermediate phenomenological theory looks in both directions.
It can be connected to the fundamental theory at one end, and at the other it
is applied directly to the experimental data.  This is an example of the
engineering attitude.  It is a pragmatic approach that is designed specifically
for use.  It is a nonspeculative procedure.  Hypotheses that go beyond what is
relevant to the available data are avoided.

Now, when we come to realm of high-energy physics, we are in a new situation.
We do not know the underlying dynamics, the underlying fundamental theory.
That raises the question of finding the best strategy.  That is, what is the
most effective way of confronting the data, of organizing it, of learning 
lessons from results within a limited domain of experimental material?

I want to argue that we should adopt a pragmatic engineering approach.
What we should {\em not\/} do is try to begin with some fundamental theory
and calculate.  As we saw, this is not the best thing to do even when you have
a fundamental theory, and if you don't have one, it's certainly the wrong thing
to do.

Historically, relativistic quantum mechanics had proved very successful in
explaining atomic and nuclear physics until we got accelerators sufficiently
high in energy to create the strongly interacting particles, which include
particles that are highly unstable and decay through very strong forces.
The ordinary methods that had evolved up to this point were simply powerless
in the face of this new situation.  At the higher energies, particles can
be---and are---created and destroyed with high probability.

In other words, the immutability of the particle---a foundation of ordinary
physics---had disappeared.

If the immutable particle has ceased to exist as the fundamental concept in 
terms of which a situation can be described, what do we replace it with?  There
have been two different points of view about how to construct a fundamental
theory for the strong interactions.

The first---the point of view of conventional operator field theory---proposes
to replace the particle with three-dimensional space itself.  In other words,
you think of energy, momentum, electric charge, and other properties as
distributed throughout space, and of small volumes of three-dimensional space
as the things that replace particles.  These volumes are the carriers of
energy, momentum, and so on.

People, including myself, have been actively developing the field idea for many
years.  I believe that this kind of theory may be the ultimate answer, but 
please recognize that it is a speculation.  It assumes that one is indeed
able to describe physical phenomena down to arbitrarily small distance, and,
of course, that goes far beyond anything we know at the moment.  All we are
able to do experimentally as we go to higher and higher energies is to plumb to 
smaller and smaller distances, but never to zero distance.

The question is, should you, in discussing the phenomena that are presently
known, make use of a speculative hypothesis like operator field theory? Can
we not discuss particle phenomenology and handle the correlations and 
organization of data without becoming involved in a speculative theory?
In operator field theory you cannot separate particle phenomena from
speculations about the structure of particles.  The operators of quantum-%
mechanical field theory conceptually mix these together.

To be able to discuss anything from the operator-field-theory point of view,
you must accept its fundamental hypothesis.  You have to accept a speculation
about how particles are constructed before you can begin to discuss how 
particles interact with each other.

Historically, this has proved to be a very difficult program to apply, and 
people have, of course, been anxious to deal directly with the experimental
data, and so there has been a reaction.  The extreme reaction to operator field
theory is to insist that there is nothing more fundamental than particles and
that, when you have a number of particles colliding with each other and the 
number of particles ceases to be constant,  all you can do is correlate what
comes into a collision with what goes out, and cease to describe in detail
what is happening during the course of the collision.

This point of view is called $S$-matrix theory.  The quantitative description
is in terms of a scattering matrix that connects the outgoing state with the
incoming state.  In this theory the particles are basic and cannot be analyzed.
Then, of course, the question comes up: what distinguishes the particular
set of particles that do exist from any other conceivable set?

The only answer that has been suggested is that the observed particles exist
as a consequence of self-consistency.  Given a certain set of particles, other
particles can be formed as aggregates or composites of these.  On the other 
hand, if particles are unanalyzable, then this should not be a new set of 
particles, but the very particles themselves.

That is the second idea, but I beg you to appreciate that it is also a
speculation.  We do not know for a fact that our present inability to describe
things in terms of something more fundamental than particles reflects an
intrinsic impossibility.

So these are the two polarized extremes in the search for a fundamental 
theory---the operator-field-theory point of view and the $S$-matrix point of
view.  Now my reaction to all of this is to ask again why we must speculate,
since the probability of falling on the right speculation is very small.

Can we not separate the theoretical problem of describing the properties of
these particles from speculations about their inner structure?  Can we not set
aside the speculation of whether particles are made from operator fields or
are made from nothing but themselves, and find an intermediate theory, a
phenomenological theory that directly confronts the data, but that is still
a creative theory?

This theory should be sufficiently flexible so that it can make contact with a
future, more fundamental theory of the structure of particles, if indeed
any more fundamental theory ever appears.  This is the line of reasoning
that led me to consider the theoretical problem for high-energy physics
from an engineering point of view.  Clearly I have some ideas in mind about
how to carry out such a program, and I would like to give you an enormously
simplified account of them.

We want to eliminate speculation and take a pragmatic approach.  We are not 
going to say that particles are made out of fields, or that particles sustain
each other.  We are simply going to say that particles are what the 
experimentalists say they are.  But we will construct a theory and not an
experimenter's manual in that we will look at realistic experimental procedures
and pick out their essence through idealizations.

There is one characteristic that the high-energy particles have in common---they
must be created.  Through the act of creation, we can define what we mean by
a particle.  How, in fact, do you create a particle?  By a collision.
The experimentalist arranges for a beam of particles to fall on a target.  In
the center-of-mass system, the target is just another beam, so two beams of
particles are colliding.  Out of the collision, the particle that we are
interested in may be produced.

We say that it is a particle rather than a random bump on an excitation curve 
because its properties are reproducible.  We still recognize the same particle
event though we vary a number of experimental parameters, such as energy, 
angles, and the kind of reaction.  The properties of the particle in question
remain the same---it has the same mass, the same spin, the same charge.

These criteria can be applied to an object that may last for only $10^{-24}$ 
sec---which decays even before it gets out of the nucleus in which it was
created.  Nevertheless, it is still useful to call this kind of object a
particle because it possesses essentially all of the characteristics that we
associate with the particle concept.

What is significant is that within a somewhat controllable region of space and
time, the properties characteristic of the particle have become transformed into
the particle itself.  The other particles in the collision are there only to 
supply the net balance of properties.  They are idealized as the {\em source\/}
of the particle.

This is our new theoretical construct.  We introduce a quantitative 
description of the particle source in terms of a source function, $S(x)$,
where $x$ refers to the space and time coordinates.  This function indicates
that, to some extent, we can control the region the particle comes from.

But we do not have to claim that we can make the source arbitrarily small as in
operator field theory.  We leave this question open, to be tested by future
experiment.

A particular source may be more effective in making particles that go in one
direction rather than another, so there must be another degree of control
expressed by a source function of momentum, $S(p)$.  But from quantum
mechanics we know that the dimension of the system and the degree of
directionality are closely related.  The smaller the system, the less
directional it can be.  And relativistic mechanics is incorporated from the 
very beginning in that the energy and the momentum are related to its mass
in the usual relativistic way.

Now the experimenter's job only begins with the production of a beam.  At the 
other end, he must detect the particles.  What is detection?  Unstable
particles eventually decay, and the decay process is a detection device.
More generally, any detection device can be regarded as a kind of collision
that annihilates the particle and hands its properties on in a more usable
form.  Thus the source concept can again be introduced as an abstraction of
an annihilation collision, with the source acting negatively, as a sink.

We now have a complete theoretical picture of any physical situation, in which
sources are used to create the initial particle of interest from the vacuum
state, and sources are used to detect the final particles resulting from some
interaction, thus returning to the vacuum state.

[Schwinger then wrote down an expression that describes the probability
amplitude that the vacuum state before sources act remains the vacuum state
after sources act, the vacuum persistence amplitude.]  The basic things that
appear in this expression are the source functions and space-time functions
that represent the state into which the particle is emitted and from which
it is absorbed, thus describing the intermediate propagation of the particle.

This simple expression can be generalized to apply to particles that have
charge, spin, etc., and to situations where more than one particle is present
at a time.  Interactions between particles are described in terms containing
more than two sources.

Our starting point accepts particles as fundamental---we use sources to
identify the particles and to incorporate a simplified view of dynamics.
From that we evolve a more complete dynamical theory in which we combine 
simple source arrangements like building blocks to produce descriptions of
situations that can in principle be as complex as we want.

A first test of this approach would be to see if we can reproduce the results
of some well-established theory such as quantum-electrodynamics.  What is the
starting point in this attack on electrodynamics?  It is the photon, a particle
that we know has certain striking properties such as zero rest mass and 
helicity 1. So we must include all these aspects of the photon in the picture,
and describe how photons are emitted and absorbed.  In consequence, the source
must be a vector, and it must be divergenceless.

This approach leads us to something resembling a vector potential, and when 
we ask what differential equations it satisfies we find they are Maxwell's
equations.  We start with the concept of the source as primary and are led
to Maxwell's differential equations as derived concepts.

The description of interactions follows the tentative procedures of life in
the real world.  The theory is not stated once and for all.  It begins with
simple phenomena---for example, accelerated charges radiate.  It then 
extrapolates that information outside its domain, predicts more complicated
phenomena, and awaits the test of experiment.  We do not begin with a final
description of, say, electron scattering.  We extrapolate to it from more
elementary situations, and this is still not the final description.

As the theory develops and becomes more encompassing, we go back to refine the
description of the scattering process and obtain a more quantitative
account of it.  This is the concept of an interaction skeleton.  The
process is there but it is not finally described to start with, its
existence is merely recognized.  This simplified reconstruction of 
electrodynamics is completely successful.

To indicate the wide sweep of the new approach, I mention that classical
gravitation theory (Einstein) can be reconstructed and simplified in a similar
way by beginning with the quantum relativistic properties of the basic 
particle, the graviton, although here indirect evidence for its properties
must be adduced.

But the real proving ground for source theory comes from the domain for
which it was invented, strong interactions.  The starting point is experimental
information at low energies.  The tentative extrapolations are toward higher
energies.  The method is quite elementary compared to other current 
techniques.  The successful correlations that have been obtained emphasize
the completely phenomenological nature of our present knowledge about particles
and refute attempts to lend fundamental credence to this or that particle
model.

A more fundamental theory may come into being one day, but it will be the
outcome of continued experimental probing to higher energies, and will 
doubtless involve theoretical concepts that are now only dimly seen.  But
that day will be greatly speeded if the flood of experimental results is
organized and analyzed with the aid of a theory that does not have built into
it a preconception about the very question that is being attacked.  This
theory is source theory.
\end{quotation}

\subsection{The Impact of Source Theory}
Robert Finkelstein has offered a perceptive discussion of Schwinger's source
theory program:
\begin{quote}
In comparing operator field theory with source theory Julian revealed his
political orientation when he described operator field theory as a trickle
down theory (after a failed economic theory)---since it descends from implicit
assumptions about unknown phenomena at inaccessible and very high energies to
make predictions at lower energies.  Source theory on the other hand he
described as anabatic (as in Xenophon's Anabasis) by which he meant that it
began with solid knowledge about known phenomena at accessible energies to make
predictions about physical phenomena at higher energies.  
Although source theory
was new, it did not represent a complete break with the past but rather was a
natural evolution of Julian's work with operator Green's functions.  His
trilogy on source theory is not only a stunning display of Julian's power
as an analyst but it is also totally in the spirit of the modest scientific
goals he had set in his QED work and which had guided him earlier as a
nuclear phenomenologist.\citex{fink}\end{quote}

But the new approach was not well received.  In part this was because
the times were changing; within a few years, 't Hooft\citex{thooft}
 would establish
the renormalizability of the Glashow-Weinberg-Salam $SU(2)\times U(1)$
electroweak model, and field theory was seen by all to be viable again.
With the discovery of asymptotic freedom in 1974,\citex{af}
 a non-Abelian gauge theory of strong interactions,
quantum chromodynamics, which was proposed somewhat earlier,\citex{QCD}
was promptly accepted  by nearly
everyone.  An alternative to conventional field theory did not seem to
be required after all. Schwinger's insistence on a clean break with the
past, and his rejection of ``rules'' as opposed to learning through
serving as an ``apprentice,'' did not encourage conversions.

Already before the source theory revolution,
 Schwinger felt a growing sense of unease with his colleagues at Harvard.
But the chief reason Schwinger left Harvard for UCLA was health related.
Formerly overweight and inactive,
he had become health conscious upon the premature death of Wolfgang Pauli
in 1958.  (Ironically, both died of pancreatic cancer.)
He had been fond of tennis from his youth, had discovered
skiing in 1960, and now his doctor was recommending a daily swim for his
health.  So he listened favorably to the entreaties of David Saxon,
his closest colleague at the Radiation Lab during the war, who
for years had been trying to induce him to come to UCLA. Very much against
his wife's wishes, he made the move in 1971.  He brought along his three
senior students at the time, Lester DeRaad, Jr., Wu-yang Tsai, and the
present author,
 who became long-term ``assistants'' at UCLA.
He and Saxon expected, as in the early days at Harvard, that
 students would flock
to UCLA to work with him; but they did not.  Schwinger was no longer the
center of theoretical physics.

This is not to say that his little group at UCLA did not make an heroic
attempt to establish a source-theory presence.  Schwinger remained a
gifted innovator and an awesome calculator.  He wrote 2-1/2 volumes of
an exhaustive treatise on source theory, 
{\it Particles, Sources, and Fields\/},\citex{153} 
devoted primarily to the reconstruction of quantum
electrodynamics in the new language; unfortunately, he abandoned the
project when it came time to deal with strong interactions, in part because
he became too busy writing papers on an ``anti-parton'' interpretation of
the results of deep-inelastic scattering experiments.\citex{167} 
He made some significant contributions
to the theory of magnetic charge; particularly noteworthy was his introduction
of dyons.\citex{147}  He reinvigorated proper-time methods
of calculating processes in strong-field electrodynamics;\citex{156}
and he made some major contributions to the theory of the Casimir effect,
which are still having repercussions.\citex{174}
But it was clear he was reacting, not leading, as witnessed by his
quite pretty paper on the ``Multispinor Basis of Fermi-Bose 
Transformation,''\citex{190} in which he kicked himself for not discovering 
supersymmetry, following a command private performance by Stanley Deser on
supergravity.

\section{The Axial-Vector Anomaly and Schwinger's Departure from 
Particle Physics}
\label{avanomaly}
In 1980 Schwinger gave a seminar at MIT that marked his last scientific
visit to the East Coast,\footnote{This does not count a talk he gave at
MIT in 1991 in honor of birthdays of two of his students, where he gave
a ``progress report'' on his work on cold fusion and sonoluminescence,
excerpts of which is given in Ref.~\cite{mm}.}
 and caused him to abandon his attempt to 
influence the development of high-energy theory with his source theory
revolution.  The talk was on a subject that he largely started in his
famous ``Gauge Invariance and Vacuum Polarization'' paper,\citex{64}
the triangle or axial-vector anomaly.  In its simplest and basic manifestation,
this ``anomaly'' describes how the neutral pion decays into two photons.
The pion coupling could be regarded as occurring either through a pseudoscalar
or an axial vector coupling, which formally appeared be equivalent, but
calculations in the 1940s gave discrepant answers.  Schwinger resolved this
issue in 1950 by showing that the two theories were indeed equivalent provided
that proper care (gauge-invariance) was used, and that the formal result
was modified by an additional term.  Problem solved, and it was then forgotten
for the next 18 years.  In the late 1960s Adler, Bell, and Jackiw rediscovered
this solution,\citex{abj} but the language was a bit different.  The
extra term Schwinger had found was now called an anomaly, but the form of the
equations, and the prediction for the decay of the pion, were identical.  In
fact, at first it is apparent that Adler, Bell, and Jackiw were unaware of
Schwinger's much earlier result, and it was the addition of Ken Johnson
(one of Schwinger's many brilliant
students) into the collaboration that corrected
the historical record.\citex{jackiwjohnson}

Shortly thereafter, Adler and Bardeen proved the ``nonrenormalization''
theorem,\citex{adlerbardeen} that the anomaly is exact, and is not corrected
by higher-order quantum effects.  This is in contrast to most physical
phenomena, such as the anomalous magnetic moment of the electron, which is
subject to corrections in all orders of perturbation theory in the strength
of the electromagnetic coupling, the fine structure constant.  This seemed
surprising to Schwinger, so he suggested to his postdocs at UCLA that
they work this out independently, and they did, publishing two papers in
1972,\citex{dmt} in which they showed, using two independent methods,
that there was indeed such a correction
in higher order.  However, Adler, who was the reviewer of these papers
forced them to tone down their conclusion, and to point out that the
result depends on the physical point at which the renormalization is
carried out.  Nonrenormalization indeed can be achieved by renormalization
at an unphysical point, which may be acceptable for the use of the
theorem in establishing renormalizability of gauge theories,
its chief application, but it is
nevertheless true that physical processes such as the original process
of pion decay receives higher-order corrections.

As was typical, Schwinger apparently took no notice of this dispute at the
time.  But toward the end of the 1970s, while he was writing the third
volume of {\it Particles, Sources, and Fields}, he looked at the questions
of radiative corrections to neutral pion decay and found the same result
as DeRaad, Milton, and Tsai.  He wrote an explicitly confrontational
paper on the subject, which was the basis for the above-mentioned talk
at MIT.  The paper was apparently definitively rejected, and the talk was
harshly criticized, and on the basis of these closed-minded attacks,
Schwinger left the field.  Fortunately for the record, Schwinger's paper
exists as a chapter in the finally published third volume of {\it Particles,
Sources, and Fields.}

However, the controversy lives on.  In 2004 Steve Adler wrote a historical
perspective on his work on the axial-vector anomaly.\citex{sla}
He devotes five pages of his retrospective to attack the work of Schwinger
and his group.  He even denies that Schwinger was the first to calculate
the anomaly, in blatant disregard of the historical record.  Of course,
physical understanding had increased in the nearly two decades between
Schwinger's and Adler's papers, but to deny that Schwinger
was the first person to offer
the basis for the connection between the axial-vector and pseudoscalar
currents, and the origin of the photonic decay of the neutral pion, is
preposterous.

\section{Thomas-Fermi Atom, Cold Fusion, and Sonoluminescence}

When the last of his Harvard postdocs left UCLA in 1979, 
and the flap over the axial-vector anomaly ensued, Schwinger abandoned
high-energy physics altogether.  In 1980,
after teaching a quantum mechanics course
(a not-unusual sequence of events), Schwinger began a series
of papers on the Thomas-Fermi model of atoms.\citex{192}  He soon
hired Berthold-Georg Englert, replacing Milton as a postdoc, to help with
the elaborate
calculations. This endeavor lasted until 1985.
It is interesting
that this work not only is regarded as important in its own right by
atomic physicists, but has led to some significant results in
mathematics.  A long series of substantial papers by C. Fefferman
and L. Seco\citex{feffsec}
 has been devoted to proving his conjecture about the
atomic number dependence of the ground state energy of large atoms.
As Seth Putterman has remarked, it is likely that, of all the work
that Schwinger accomplished at UCLA, his work on the statistical atom
will prove the most important.\citex{seth}

Following the Thomas-Fermi work, Schwinger continued to collaborate
with Englert, and with Marlan Scully, on the question of spin coherence.  
If an atomic beam is separated into sub-beams by a Stern-Gerlach apparatus, 
is it possible to reunite the beams?  Scully had argued that it might
be possible, but Julian was skeptical; the result was three joint
papers, entitled ``Is Spin Coherence Like Humpty Dumpty?'', which bore
out Julian's intuition of the impossibility of beating the effects of
quantum entanglement.\citex{humptydumpty}

In March  1989 began one of the most curious episodes in physical
science in the last century, one that initially attracted great interest
among the scientific as well as the lay community, but which rapidly
appeared to be a characteristic example of ``pathological
science.''\footnote{This term was coined in 1953 by Irving Langmuir,
who gave a celebrated lecture at General Electric's Knolls Atomic Power
Laboratory (transcribed from a disc recording by Robert Hall)
 on the phenomenon wherein reputable
scientists are led to believe  that an effect, just at the edge of visibility,
is real, even though, as precision increases, the effect remains
marginal.  The scientist becomes self-deluded, going to great lengths
to convince one and all that the remarkable effect is there just on
the margins of what can be measured. Great accuracy is claimed nevertheless,
and fantastic, {\it ad hoc}, theories are invented to explain the effect.
 Examples include N-rays, the
Allison effect, flying saucers, and ESP. It was not a coincidence that {\it
Physics Today\/} published the article, without comment, in the fall of
1989.\citex{langmuir}}
The effect to which we refer was the announcement by B.  S. Pons and
M. Fleischmann\citex{pf}
 of the discovery of cold fusion.  That is, they claimed
that nuclear energy, in the form of heat, was released in a table-top
experiment, involving a palladium cathode electrolyzing heavy water.

So it was a shock to most 
physicists\footnote{However, a few other
eminent physicists spoke favorably of the possibility of cold fusion,
notably Edward Teller and Willis Lamb, who published three articles
in the {\it Proceedings of the U.S. National Academy of Sciences\/}
 on the subject.}
 when Schwinger began speaking and
writing about cold fusion, suggesting that the experiments of Pons and
Fleischmann were valid, and that the palladium lattice played a crucial
role. In one of his later lectures on the subject in Salt Lake City,
Schwinger recalled, ``Apart from a brief period of apostasy, when I echoed
the conventional
wisdom that atomic and nuclear energy scales are much too disparate, I have
retained my belief in the importance of the
lattice.''\citex{archive}
 His first publication on the subject was submitted to {\it Physical
Review Letters}, but was roundly rejected, in a manner that Schwinger
considered deeply insulting.  In protest, he resigned as a member
(he was, of course, a fellow) of the American Physical Society, of
which {\it Physical Review Letters\/} is the most prestigious journal.
(At first he intended merely to withdraw the paper from
{\it PRL}, and his fellowship, but then he felt compelled to
respond to the referees' comments: One comment was something to the
effect that no nuclear physicist could believe such an effect, to
which Julian angrily retorted, ``I am a nuclear physicist!''\citex{archive})
 In this letter to the editor (G. Wells) in which
he withdrew the paper and resigned from the American Physical Society,
he also called for the removal of
the source theory index category the APS journals
used: ``Incidentally, the PACS entry (1987) 11.10.mn can be
deleted.  There will be no further occasion to use
it.';\citex{englertnote,archive}
 A rather striking
act of hubris: If he couldn't publish source theory, neither could anybody
else.  But the {\it Physical Review\/} obliged.  (Unfortunately, 
Schwinger failed to realize that the PACS
index system has become the predominant system for physics journals
worldwide, a reflection of the premier status of the APS journals. So
he largely spited his own contributions.)
 Not wishing to use any other APS venue, he turned to his friend and colleague,
Berthold Englert, who arranged that ``Cold Fusion: A Hypothesis''
 be published in the {\it Zeitschrift
f\"ur Naturforschung}, where it appeared in October of that year.\citex{213}
Schwinger then went on to write three substantial papers, entitled
``Nuclear Energy in an Atomic Lattice I, II, III,'' to flesh out
these ideas.\citex{englertx,archive}
The first was published in the {\it Zeitschrift f\"ur Physik D},\citex{214}
where it was accepted in spite of negative
reviews,\citex{archive}
but directly preceded by an editorial note, disclaiming any responsibility
for the the paper on the part of the journal.  They subsequently refused
to publish the remaining papers.

\setcounter{footnote}{0}
Schwinger's last physics endeavor marked a return to the Casimir effect,
of which he had been enamored nearly two decades earlier.  It was sparked by
the remarkable discovery of single-bubble sonoluminescence,
in which a small bubble of air in water, driven by a strong acoustic
standing wave, undergoes a stable cycle of collapse and re-expansion;
at minimum radius an intense flash of light, consisting of a million
optical photons, is emitted.   It was
not coincidental that the leading laboratory investigating this phenomenon
was, and is, at UCLA, led by erstwhile theorist Seth Putterman, long
a friend and confidant.  Putterman and Schwinger shared many interests
in common, including appreciation of fine wines, and they shared a similar
iconoclastic view of the decline of physics.  So, of course, Schwinger
heard about this remarkable phenomenon from the horse's mouth, and was
greatly intrigued.\footnote{For a review of the phenomena, and a
detailed evaluation of various theoretical explanations, see
Ref.~\cite{sonorev}.}

Schwinger immediately had the idea that
a dynamical version of the Casimir effect might play a key role.
He saw parallels between cold fusion and sonoluminescence in that both
deal with seemingly incommensurate energy scales, and both depend
significantly on nonlinear effects.  Since by the early 1990s, cold
fusion was largely discredited, he put his efforts to understanding
sonoluminescence, which undoubtedly does exist.  Unfortunately neither
Schwinger, nor anyone subsequently, was able to get very far
with dynamical zero-point phenomena; he largely contented himself with
an adiabatic approximation based on static Casimir energies; and was
able to obtain sufficient energy only because he retained the ``bulk
energy,'' which most now believe is unobservable, being subsumed
in a renormalization
of bulk material properties.  His work on the subject appeared as
a series of short papers in the PNAS, the last appearing\citex{jssono}
shortly after his death in June 1994.

\section{Conclusion}
It is impossible to do justice in a few words
to the impact of Julian Schwinger on physical thought in the 20th Century.
He revolutionized fields from nuclear physics to many body theory,
first successfully formulated renormalized quantum electrodynamics,
developed the most powerful functional formulation of quantum field theory,
and proposed new ways of looking at quantum mechanics, angular momentum
theory, and quantum fluctuations.  His legacy includes ``theoretical tools''
such as the proper-time method, the quantum action principle, and effective
action techniques.  Not only is he responsible for formulations bearing his
name: the Rarita-Schwinger equation, the Lippmann-Schwinger equation, the
Tomonaga-Schwinger equation, the Dyson-Schwinger equation,
 the Schwinger mechanism, and so forth, but
some attributed to others, or known anonymously:
Feynman parameters, the Bethe-Salpeter equation,
coherent states, Euclidean field theory; the list goes on and on.  His legacy
of nearly 80 Ph.D. students, including four Nobel laureates, lives on.
It is impossible to imagine what physics would be like in the 21st century
without the contributions of Julian Schwinger, a very private
yet wonderful human being.  It is most gratifying that a dozen years after his
death, recognition of his manifold influences is growing, and research projects
he initiated are still underway.

It is fitting to close this retrospective with Schwinger's own words, delivered
some six months before his final illness, when he received an honorary
degree from the University of Nottingham.\citex{ng}\footnote{This brief
acceptance speech was followed by a brilliant lecture on the influence of
George Green on Schwinger's work.\citex{ng}}
\begin{quotation}
The Degree Ceremony is a modern version of a medieval rite that seemed to
confer a kind of priesthood upon its recipients, thereby excluding all others
from its inner circle.  But that will not do for today.  Science, with its
offshoot of Technology, has an overwhelming impact upon all of us.  The recent
events at Wimbledon invite me to a somewhat outrageous analogy.  
Very few of us,
indeed, are qualified to step onto centre court.  Yet thousands of spectators
gain great pleasure from watching these talented specialists perform.  
Something
similar should be, but generally is not, true for the relationship between
the practitioners of Science and the general public.  This is much more serious
than not knowing the difference between 30 all and deuce.  Science, on a big
scale, is inevitably intertwined with politics.  And politicians have little
practice in distinguishing between, say common law and Newton's law.  It is
a suitably educated public that must step into the breach.  This has been
underlined lately by Minister Waldegrave's cry for someone to educate him
about the properties of the Higgs boson, to be rewarded with a bottle of
champagne.  Any member of the educated public could have told him that the
cited particle is an artifact of a particular theoretical speculation, and the
real challenge is to enter uncharted waters to see what is there.  The failure
to do this will inevitably put an end to Science.  A society that turn in on
itself has sown the seeds of its own demise.  Early in the 16th century,
powerful China had sea-going vessels exploring to the west.  Then a signal 
came from new masters to return and destroy the ships.  It was in those years
that Portuguese sailors entered the Indian Ocean.  The outcome was 400 years
of dominance of the East by the West.

There are other threats to Science.  A recent bestseller in England, {\it 
Understanding the present}, has the subtitle {\it Science and the soul of
Modern Man}.  I shall only touch on the writer's views toward quantum
mechanics, surely the greatest intellectual discovery of the 20th century.
First, he complains that the new physics of quantum mechanics tosses classical
physics in the trash bin.  This I would dismiss as mere technical ignorance;
the manner in which classical and quantum mechanics blend into each other has
long been established.  Second, the author is upset that its theories can't
be understood by anyone not mathematically sophisticated and so must be
accepted by most people on faith.  He is, in short, saying that there is a
priesthood.  Against this I pose my own experience in presenting the basic
concepts of quantum mechanics to a class of American high school students.
They understood it; they loved it.  And I used no more than a bit of algebra,
a bit of geometry.  So: catch them young; educate them properly; and there are
no mysteries, no priests.  It all comes down to a properly educated public.
\end{quotation}

\section*{Acknowledgements}
I am greatful to many colleagues for the interviews and conversations
granted me in writing about Julian Schwinger.  I am particularly grateful
to Robert Finkelstein and Edward Gerjuoy for conversations in the past few
months.  Again I must thank Charlotte Brown, Curator of Special Collections
at UCLA, for making the Schwinger archives available to me on many occasions.
My research over the year, not primarily historical, has been funded by
grants from the US Department of Energy and the US National Science Foundation.
I dedicate this memoir to Julian's widow, Clarice.

\end{document}